\documentclass[pdflatex,fleqn]{sn-jnl}
\usepackage{savesym}
\savesymbol{lneq}
\let\origtheorem\theorem
\let\origproof\proof
\usepackage{z-eves}
\let\theorem\origtheorem
\let\proof\origproof
\restoresymbol{TXF}{lneq}
\usepackage[utf8]{inputenc}
\usepackage{url}
\usepackage{comment}
\usepackage{soul}
\usepackage{xspace}
\usepackage{xcolor}
\usepackage{booktabs}
\usepackage{colortbl}
\usepackage{tabularx}
\usepackage{comment}
\usepackage{algorithm}
\usepackage{algpseudocode}
\usepackage{fancyvrb}
\usepackage{paralist}
\usepackage[inference]{semantic}

\newtheorem{remark}{Remark}
\newtheorem{lemma}{Lemma}
\newtheorem{example}{Example}

\newcommand{\setlog}{$\{log\}$\xspace}
\newcommand{\SET}{\mathcal{HFS}}
\newcommand{\BR}{\mathcal{BR}}
\newcommand{\RIS}{\mathcal{RIS}}
\newcommand{\LCARD}{\mathcal{CARD}}
\newcommand{\LINT}{\mathcal{INTV}}
\newcommand{\LIA}{\mathcal{LIA}}

\renewcommand{\Cup}{\mathtt{un}}
\renewcommand{\Cap}{\mathtt{inters}}
\newcommand{\In}{\mathbin{\mathtt{in}}}
\newcommand{\Dom}{\mathtt{dom}}
\newcommand{\Subseteq}{\mathtt{subset}}
\newcommand{\Comp}{\mathtt{comp}}
\newcommand{\ApplyTo}{\mathtt{applyTo}}
\newcommand{\Pfun}{\mathtt{pfun}}

\newcommand{\Nin}{\mathbin{\mathtt{nin}}}
\newcommand{\Ris}{\mathtt{ris}}

\newcommand{\Cp}{\mathtt{cp}}

\newcommand{\Int}{\mathtt{int}}
\newcommand{\Str}{\mathtt{str}}
\newcommand{\Etype}{\mathtt{enum}}
\newcommand{\Stype}{\mathtt{set}}
\newcommand{\Rtype}{\mathtt{rel}}
\newcommand{\Sumtype}{\mathtt{sum}}
\newcommand{\Dec}{\mathtt{dec}}

\newcommand{\Or}{\mathbin{\mathtt{or}}}
\newcommand{\Neg}{\mathtt{neg}}
\renewcommand{\Forall}{\mathtt{foreach}}
\renewcommand{\Exists}{\mathtt{exists}}
\newcommand{\Implies}{\mathbin{\mathtt{implies}}}
\renewcommand{\false}{\mathtt{false}}

\newcommand{\step}[1]{\mathbin{\lower0.55ex\hbox{$\lhook\joinrel\xrightarrow{#1}$}}}

\newcommand{\AndroidState}{\mathsf{AndroidST}}

\newcommand{\Action}{\mathsf{Action}}
\newlength{\bcextramargin}
\newenvironment{changemargin}[2]{\begin{list}{}{%
\setlength{\topsep}{0pt}%
\setlength{\leftmargin}{0pt}%
\setlength{\rightmargin}{0pt}%
\setlength{\listparindent}{\parindent}%
\setlength{\itemindent}{\parindent}%
\setlength{\parsep}{0pt plus 1pt}%
\addtolength{\leftmargin}{#1}%
\addtolength{\rightmargin}{#2}%
}\item }{\end{list}} 

\newcommand{\actdefsection}[1]{
\begin{changemargin}{-\bcextramargin}{0pt}
\vspace{1ex}
\noindent
\textbf{{#1}}
\end{changemargin}
}

\newenvironment{absolutelynopagebreak}
  {\par\nobreak\vfil\penalty0\vfilneg
   \vtop\bgroup}
  {\par\xdef\tpd{\the\prevdepth}\egroup
   \prevdepth=\tpd}

\begin{document}

\title[A Verified Prototype of the Android Permissions System]%
  {An Automatically Verified Prototype of the Android Permissions System}

\author*[1]{\fnm{Maximiliano} \sur{Cristi\'a}}\email{cristia@cifasis-conicet.gov.ar}
\affil*[1]{Universidad Nacional de Rosario and CIFASIS,
Argentina}
\equalcont{All authors contributed equally to this work}

\author[2]{\fnm{Guido} \sur{De Luca}}\email{gdeluca@dcc.fceia.unr.edu.ar}
\affil[2]{Universidad Nacional de Rosario, Argentina}
\equalcont{All authors contributed equally to this work}

\author[3]{\fnm{Carlos} \sur{Luna}}\email{cluna@fing.edu.uy}
\affil[3]{Universidad de la Rep\'ublica, Uruguay}
\equalcont{All authors contributed equally to this work}

\abstract{%
In a previous work De Luca and Luna presented formal specifications
of idealized formulations of the permission model
of Android in the Coq proof assistant. This formal development is about
23 KLOC of Coq code, including proofs.
This work aims at showing that \setlog (`setlog')---a satisfiability solver and a constraint logic programming language---can be used as an effective automated prover
for the class of proofs that must be discharged in the formal verification of systems
such as the one carried out by De Luca and Luna.
We show how the Coq model is encoded in \setlog and how automated proofs are performed. The resulting \setlog model is an automatically verified executable prototype of the Android permissions system. Detailed data
on the empirical evaluation resulting after executing all the proofs in \setlog is provided. The integration of Coq and \setlog as to provide a framework featuring
automated proof and prototype generation is discussed.
}


\maketitle

\section{Introduction}\label{sec:introduction}
Mobile devices have become an integral part of how people perform tasks in their work and personal lives. The benefits of using mobile devices, however, are sometimes offset by increassing security risks.

Android \cite{AndroidProy} is an open platform for mobile devices---developed by the Open Handset Alliance led by Google, Inc---that captures more than 85\% of the total market-share \cite{reporteIDC}.
Android offers many critical applications in terms of privacy. In order to provide their users the security they expect, Android relies on a multi-party consensus system where user, OS and application must be all in favor of performing a task.
Android embodies security mechanisms at both OS and application level.
In particular, Android behaves like a multi-threaded Linux system; its security model is similar to that of a multi-user server. Application-level access control is implemented by an Inter-Component Communication reference monitor that enforces mandatory access control policies regulating access among applications and components.
Application security is built primarily upon a system of permissions, which specify restrictions on the operations a particular process can perform.

In previous works two of the authors
present formal specifications in the Coq proof assistant \cite{CoqRM,DBLP:series/txtcs/BertotC04} of idealized formulations of different versions of the permission model of Android \cite{DBLP:conf/ictac/BetarteCLR15,DBLP:journals/cuza/BetarteCLR16,DBLP:conf/lopstr/BetarteCGL17}.
These works formulate and demonstrate, in a non-automatic way, a set of properties of the Android security model. In particular, De Luca and Luna \cite{DBLP:conf/types/Luca020} formalize the Android permissions system introduced in versions Nougat, Oreo, Pie and 10. The formal development is about 23 KLOC of Coq code, proofs included.

The present work aims at showing that \setlog \cite{setlog}---a satisfiability solver and a constraint logic programming language---can be used as an effective automated prover for the class of proofs that must be discharged in the formal verification of systems such as the one carried out by De Luca and Luna. In this way much of the manual, expert work needed to prove properties in Coq can be avoided. In particular, Cristi\'a, assisted by De Luca and Luna, encoded in \setlog the Coq model developed by his coauthors. Hence, this paper shows in detail how different elements and properties of the Coq model can be encoded and proved in \setlog. The resulting \setlog model is an executable prototype. Additionally, detailed data on the empirical evaluation resulting after executing all the proofs in \setlog is provided. Finally, the integration of Coq and \setlog as to provide a framework with automated proof and prototype generation is briefly discussed.

To the best of our knowledge this is the first time that Coq and \setlog are put together for the verification of a critical system.

\subparagraph{Organization of the paper}
The rest of the paper is organized as follows.
Section \ref{background} provides some background on the Android permission system, its Coq model and \setlog.
Next, Section \ref{encoding} shows how different elements of the Coq model are encoded in the \setlog language, whereas
Section \ref{prototype} shows how the properties proved true of the Coq model can be encoded as \setlog satisfiability queries.
Section \ref{quantrep} provides detailed data on the empirical evaluation resulting after executing all the \setlog queries described.
Section \ref{relwork} considers related work and finally, Section \ref{concl} concludes with a summary of our contributions and directions for future work.

\section{\label{background}Background}
This section contains some background on the Android permissions system (\ref{android}); the Coq model developed by De Luca and Luna (\ref{coq}); and the \setlog constraint solver (\ref{setlog}). Readers may skip any of these subsections if they are familiarized with those topics.

\subsection{\label{android}The Android Permissions System}

The Android security model 
takes advantage of the fact that the entire platform is built upon a Linux
kernel and relies primarily on a user-based protection to identify and isolate the resources from
each application. In other words, each Android application will have a unique user ID (UID), with
restricted permissions, and will run its own process in a private virtual machine. This means that
by default, applications will not be able to interact with each other and they will also have limited
access to OS resources. 

In the Android permissions system, every
resource is protected by unique tags or names, called \textit{permissions}, that applications must have been granted before being
able to interact with the resource.
Permissions can be defined by applications, for the sake of self-protection; and are predefined by
Android so applications can get access to system resources, such as the camera or other kind of
sensors. Applications must list the permissions they need in the so-called  application \textit{manifest}.
Permissions are also associated to a \textit{protection level}, depending on how critical the
resource that they are securing is. The protection level also determines if the permission will be
automatically granted upon installation or if user consent will be required at runtime. Android defines three protection levels: \emph{normal}, \emph{signature} and \emph{dangerous}.

Permissions related to the same
device capability, are grouped into \textit{permissions groups}. For example, reading and editing contacts, are
two different permissions. However, when an application requests one of these permissions
\emph{for the first time}, the user will be asked to authorize the ``contacts group''.
If the user accepts, then the behavior goes as follows:
\begin{enumerate}
    \item The requested permission is granted and the system is notified that the user authorized
          the ``contacts group''.
    \item As soon as the application requests another permission from the same group, the system
          will automatically grant it, without informing the user.
\end{enumerate}

\subsection{\label{coq}A Verified Coq Model}

As we have said, the Android permissions system has been modeled using the Coq proof assistant \cite{DBLP:conf/ictac/BetarteCLR15,DBLP:journals/cuza/BetarteCLR16,DBLP:conf/lopstr/BetarteCGL17,DBLP:conf/types/Luca020}. Coq is a formal proof management system based on higher
order logic that allows to write formal specifications and interactively generate machine-checked
proofs of theorems. It also provides a (dependently typed) functional programming language that can
be used to write executable algorithms. The Coq environment also provides program
extraction towards languages like Ocaml and Haskell for execution of certified algorithms
\cite{letouzey04}. Proofs in Coq are essentially manual and interactive.

The Coq model of the Android permissions system follows the structure of an abstract state machine. In this sense, the model defines a tuple representing the set of states of that machine (Section \ref{sec:model:stateI}) and a number of state transitions (called operations) given by means of their pre- and post-conditions (Section \ref{sec:actions}). The state of the machine contains both dynamic and static
data of the system. In turn, the operations correspond to the actions that either the user, applications or Android can perform concerning the permission system. 
The security properties that have been proved true of the the model are discussed in Section \ref{coq:properties}. 

The source code of the Coq model and the proof scripts can be found in a Github repository \cite{AndroidCoq:2020}. File names mentioned in this section refer to that repository.

\subsubsection{The State Space}
\label{sec:model:stateI}

In Figure \ref{f:coq:state}, we show a
fragment of Coq code corresponding to the state definition, called \texttt{System}. As can be seen, \texttt{System} is the union of \texttt{State} and \texttt{Environment}. The former contains information such as which applications are currently installed (\verb+apps+), which applications have been executed at least once (\verb+alreadyVerified+), which groups of permissions were already authorized (\verb+grantedPermGroups+) and which
permissions were granted to each application (\verb+perms+). In turn, \texttt{Environment} takes care of tracking information that will not change once each application is
installed. For example, it contains a mapping between each application and its manifest (\verb+manifest+), and a
mapping between each application and the permissions defined by it (\verb+defPerms+). In this way \texttt{System} sums up a total of 13
components that represents the whole state of the model. The definitions of more detailed types
are not included, for the sake of keeping the presentation brief. The types starting
with lowercase correspond to basic parameters modelled as \texttt{Set} in Coq, whereas those
starting with a capital letter were defined by more complex \texttt{Records}.
From now on, whenever we mention the model's states, we will be referring to elements of
type \texttt{System}, unless stated otherwise.

\begin{figure}
    \begin{Verbatim}[gobble=2,
                     numbers=left,
                     numbersep=5pt,
                     numberblanklines=false,
                     xleftmargin=5mm,fontsize=\small]
  Record System := sys {
          state: State;
          environment: Environment 
  }.
  Record State := st { 
      apps: list idApp;
      alreadyVerified: list idApp;
      grantedPermGroups : mapping idApp (list idGrp);
      perms: mapping idApp (list Perm);
      ...
  }.
  Record Environment := env {
      manifest: mapping idApp Manifest;
      defPerms: mapping idApp (list Perm);
      ...
  }.
    \end{Verbatim}
    \caption{\label{f:coq:state}State definition in Coq}
\end{figure}

\subparagraph{Valid states}
The model defines a notion of valid state that captures several well-formedness conditions. This is so to avoid reasoning over some values of \texttt{System} that make no sense
either in a real world scenario or in the Coq representation of the system. For example, the valid state predicate forces the permissions defined by
applications to be uniquely identified. In other words, if two user-defined permissions have
the same ID, then they were defined by the same application (and therefore, they are the same
permission). This Coq predicate, called \verb+notDupPerm+, is shown in Figure \ref{f:coq:notDupPerm}. Note
that without \verb+notDupPerm+, an application could get access to a permission that is protecting some
resource just by defining a permission of its own with the same ID.

\begin{figure}
    \begin{Verbatim}[gobble=2,
                     numbers=left,
                     numbersep=5pt,
                     numberblanklines=false,
                     xleftmargin=5mm,fontsize=\small]
  Definition notDupPerm : Prop := 
    forall (a a':idApp) (p p':Perm) (l l':list Perm),
      defPermsForApp a l -> defPermsForApp a' l' ->
      In p l -> In p' l' -> idP p = idP p' ->
      (p=p' /\ a=a').

  Definition defPermsForApp (a:idApp) (l:list Perm) : Prop :=
    map_apply (defPerms (environment s)) a = Value l
    \/ (exists sysapp:SysImgApp, 
          In sysapp (systemImage (environment s)) /\
          defPermsSI sysapp = l /\ idSI sysapp = a).
    \end{Verbatim}
    \caption{\label{f:coq:notDupPerm}One of the constraints defining the valid state condition}
\end{figure}

The valid state predicate also forces that every mapping in the state is a partial function. Figure \ref{f:coq:allMapsCorrect}
shows a fragment of this predicate, called \texttt{allMapsCorrect}. The fixpoint declaration \texttt{map\_correct} in line \ref{v:map_correct} is a
recursive function that returns true only if there are no repeated elements in the domain of the map. 
The definition of the valid state property is the result of conjoining 13 different
predicates, which are defined within 200 lines of code (file \texttt{Estado.v}).

\begin{figure}
    \begin{Verbatim}[gobble=2,
                     numbers=left,
                     numbersep=5pt,
                     numberblanklines=false,
                     xleftmargin=5mm,
                     commandchars=+\[\],fontsize=\small]
  Definition allMapsCorrect : Prop :=
    map_correct (manifest (environment s)) /\
    map_correct (cert (environment s)) /\
    map_correct (defPerms (environment s)) /\
    map_correct (grantedPermGroups (state s)) /\
    ...
    
  Fixpoint map_correct (mp:mapping) : Prop := +label[v:map_correct]
    match mp with
      | nil => True
      | (a::rest)=> ~(In (item_index a) (map_getKeys rest)) /\
                    map_correct rest
    end.
    \end{Verbatim}
    \caption{\label{f:coq:allMapsCorrect}Fragment of \texttt{allMapsCorrect}}
\end{figure}

\subsubsection{Operations as State Transitions}
\label{sec:actions}
The state transitions of the model, called
``actions'' or ``operations'', cover the main functionalities of the Android permission
system. As an
example, Figure~\ref{f:grantAutoCoq} shows the semantics of the operation \texttt{grantAuto}, which
is responsible for granting a permission to an application that already has an
authorization for automatically granting permissions from a certain group. This
action is executed whenever a dangerous permission is granted without
explicit consent from the user.

\begin{figure}
\begin{Verbatim}[gobble=2,
                 numbers=left,
                 numbersep=5pt,
                 numberblanklines=false,
                 xleftmargin=5mm,fontsize=\small]
  Definition grantAuto (p:Perm) (a:idApp) (s s':System) : Prop :=
    pre_grantAuto p a s /\ post_grantAuto p a s s'.

  Definition pre_grantAuto (p: Perm) (a: idApp) (s: System) : Prop :=
    (exists m:Manifest, isManifestOfApp a m s /\ In p (use m)) /\
    (isSystemPerm p \/ usrDefPerm p s) /\
    ~(exists lPerm:list Perm, 
        map_apply (perms (state s)) a = Value lPerm /\ In p lPerm) /\
    pl p = dangerous /\
    (exists (g: idGrp) (lGroup: list idGrp), 
        maybeGrp p = Some g /\
        map_apply (grantedPermGroups (state s)) a = Value lGroup /\
        In g lGroup).

  Definition post_grantAuto (p:Perm) (a:idApp) (s s':System) : Prop :=
    grantPerm a p s s' /\
    (environment s) = (environment s') /\
    (apps (state s)) = (apps (state s')) /\
    ...      (* Other state components remain the same *)

  Definition grantPerm (a:idApp) (p:Perm) (s s':System) : Prop :=
    (forall (a':idApp) (lPerm:list Perm),
        map_apply (perms (state s)) a' = Value lPerm ->
        exists lPerm':list Perm,
            map_apply (perms (state s')) a' = Value lPerm' /\
            forall p':Perm, In p' lPerm -> In p' lPerm') /\
    (forall (a':idApp) (lPerm':list Perm),
        map_apply (perms (state s')) a' = Value lPerm' ->
        exists lPerm:list Perm,
            map_apply (perms (state s)) a' = Value lPerm /\
            forall p':Perm, 
                In p' lPerm' -> ~In p' lPerm -> (a=a' /\ p=p')) /\
    (exists lPerm':list Perm,
        map_apply (perms (state s')) a = Value lPerm' /\
        In p lPerm') /\ 
    map_correct (perms (state s')).
\end{Verbatim}
\caption{\label{f:grantAutoCoq}Semantics of \texttt{grantAuto} in Coq}
\end{figure}

The precondition (\verb+pre_grantAuto+) establishes that the permission \verb+p+ is listed on the application's
manifest (and this manifest, of course, is required to exist). Regarding
\verb+p+, it is also required that it is defined either by the user or the system, that its
level is \verb+dangerous+ and that it has not been already granted to \verb+app+. Note that the latter is
required because lists (instead of sets) are used to keep track of the already
granted permissions (see more on sets and lists in Coq in Remark \ref{r:setslists} below). 
Last but not least, the precondition of this action also requires that \verb+p+ belongs to a group \verb+g+ that the user has previously authorized for automatic
permission granting.

The postcondition (\verb+post_grantAuto+) basically adds \verb+p+ to \verb+(perms a)+ in the new state \verb+s'+. This behavior is handled by the auxiliary
predicate \verb+(grantPerm a p s s')+. The rest of the components of the state remain the
same.

A non-extensive list of other actions that can be performed in the model is:
\begin{inparaenum}[i)]
    \item \texttt{grant}, which is the complementary operation of \texttt{grantAuto}, since it
    represents a permission being granted with explicit consent from the user;
    \item \texttt{revoke} or \texttt{revokeGroup}, to remove an ungrouped permission or all of the
    permission from a group, respectively;
    \item \texttt{hasPermission}, to check whether an application has a certain permission or not at
    a given moment.
\end{inparaenum}
The model supports 22 operations in total, some of them being considerably more complex than \verb+grantAuto+. The definition of the formal
semantics of these operations comprises around 1,400 lines of Coq code (file
\texttt{Semantica.v}).

We close this section with the following important observations.

\begin{remark}[State validity is invariant]\label{r:valid}
De Luca and Luna demonstrated using Coq that each operation preserves valid states.
\begin{lemma} [Validity is invariant]
    \label{lemma:valid-state-correct}
    \mbox{}\\
    $\begin{array}{l}
            \forall\ (s\ s':\AndroidState)(a:\Action),
            valid\_state(s) \land s\step{a}s' \rightarrow valid\_state(s')
        \end{array}$
\end{lemma}
\end{remark}

\begin{remark}[Sets and lists in Coq]\label{r:setslists}
In Coq sets of type \verb+T+ can be encoded as \verb+T -> Prop+, i.e. functions from \verb+T+ onto \verb+Prop+. 
An evident consequence of this choice is that the resulting specification is not executable. In particular, the program extraction mechanism provided by Coq to extract programs from specifications cannot be used in this case.
Hence, a specification encoding sets as \verb+T -> Prop+ should be refined into an specification encoding sets as lists, i.e. \verb+list T+, if a prototype is needed. In this case, the new specification must take care of repetitions and permutations. The Coq model of the Android permissions system developed by De Luca and Luna \cite{DBLP:conf/types/Luca020} was meant to produce a certified program. Then, De Luca and Luna use \verb+list T+ to encode sets.
\qed
\end{remark}

\subsubsection{\label{coq:properties}Properties of the Model}
De Luca, Luna and their colleagues
use Coq \cite{DBLP:conf/ictac/BetarteCLR15,DBLP:journals/cuza/BetarteCLR16,DBLP:conf/lopstr/BetarteCGL17,DBLP:conf/types/Luca020}  to analyze properties of the Coq model. In particular, several security properties establish that the model
provides protection against unauthorized access to sensitive resources of a device running
Android \cite{DBLP:conf/ictac/BetarteCLR15,DBLP:journals/cuza/BetarteCLR16}. 
Using the Coq specification 
several lemmas were proved showing that the Android security model meets the so-called \textit{principle of least
privilege}, i.e. that ``each application, by default, has access only to the components that it
requires to do its work and no more'' \cite{fundamentals}.

In a recent work, De Luca and Luna \cite{DBLP:conf/types/Luca020}  present and discuss some properties
about Android 10. In that work the focus is on safety-related properties concerning the changes introduced on
the later versions of Android (mainly Oreo and 10) rather than on security issues.

Furthermore, using the Coq model, De Luca and Luna precisely state the conditions that would help
preventing the exploitation of some well-known vulnerabilities of the Android
system \cite{Sbirlea2013,DBLP:conf/mobisys/ChinFGW11}, like the unauthorized monitoring of information
(\emph{eavesdropping}) or inter-application communication (\emph{intent spoofing}). They also
prove that, under certain hypotheses, these attacks cannot be carried
out \cite{DBLP:conf/ictac/BetarteCLR15,DBLP:journals/cuza/BetarteCLR16}. Some of these potentially dangerous behaviors may not be considered in the
informal documentation of the platform.

A total of 14 properties have been analyzed with Coq. Next we comment on two of them.

Figure \ref{f:cannotAutoGrantWithoutGroup} shows the Coq definition of the property stating that the system automatically grants a dangerous permission only when the user had previously authorized the permission
group that contains it.
Note that in the expression defined in lines 5-7, it is claimed that the group of the permission is not
granted to the application in the state \verb+s+. Therefore, line 8 concludes that the execution of
\texttt{grantAuto} cannot be successful.

\begin{figure}
\hspace{3mm}
\begin{minipage}{.9\textwidth}
\begin{Verbatim}[gobble=2,
                 numbers=left,
                 numbersep=5pt,
                 numberblanklines=false,fontsize=\small]
  Theorem CannotAutoGrantWithoutGroup :
    forall (s s': System) (p: Perm) (g: idGrp) (a: idApp),
      pl p = dangerous ->
      maybeGrp p = Some g ->
      ~(exists (lGroup: list idGrp),
           map_apply (grantedPermGroups (state s)) a = Value lGroup /\
           In g lGroup) ->
      ~ exec s (grantAuto p a) s' ok.
    \end{Verbatim}
    \end{minipage}
    \caption{\label{f:cannotAutoGrantWithoutGroup}Safety property about automatic granting}
\end{figure}

In turn, Figure \ref{f:execAutoGrantWithoutIndividualPerms} shows the Coq encoding of a scenario
where the system is able to automatically grant a dangerous permission to an application even though
the application has no other permission of the same group at the moment. This ``vulnerable'' state is reached when the user authorized
that group at some point, then all the permissions of that group were removed but the ability of automatic granting was still kept by the system. 

\begin{figure}
\hspace{3mm}
\begin{minipage}{.9\textwidth}
\begin{Verbatim}[gobble=2,
                 numbers=left,
                 numbersep=5pt,
                 numberblanklines=false,fontsize=\small]
    Theorem ExecAutoGrantWithoutIndividualPerms :
        exists (s: System) (p: Perm) (g: idGrp) (a: idApp),
        validstate s /\
        ~ (exists (p': Perm) (permsA: list Perm), 
              map_apply (perms (state s)) a = Value permsA /\
              In p' permsA /\ maybeGrp p' = Some g) /\
        pl p = dangerous /\
        maybeGrp p = Some g /\
        pre_grantAuto p a s.
    \end{Verbatim}
    \end{minipage}
    \caption{\label{f:execAutoGrantWithoutIndividualPerms}Edge case scenario found for the automatic
    granting feature}
\end{figure}

\subsection{\label{setlog}The \setlog Constraint Solver}

This paper aims at showing that \setlog can be used as an effective automated prover for systems such as the Android permissions system. In this way much of the manual, expert work needed to prove properties in Coq can be avoided. Consider that proving all the properties of the Coq model described in Section \ref{coq} requires 18 KLOC of Coq proof commands.

\setlog is a publicly available satisfiability solver and a declarative,
set-based, constraint-based programming language implemented in Prolog
\cite{setlog}. \setlog is deeply rooted in the work on Computable Set Theory
\cite{10.5555/92143,DBLP:series/mcs/CantoneOP01}, combined with the ideas put forward by the set-based
programming language SETL \cite{DBLP:books/daglib/0067831}.
Below we briefly introduce \setlog; more material can be found in Appendix \ref{app:setlog}.

The automated proving power of \setlog comes from the implementation of several decision procedures for different theories on the
domain of finite sets, finite set relation algebra and linear integer arithmetic
\cite{Dovier00, DBLP:journals/jar/CristiaR20, DBLP:conf/RelMiCS/CristiaR18, DBLP:journals/jar/CristiaR21a, cristia_rossi_2021, DBLP:journals/corr/abs-2105-03005, DBLP:journals/corr/abs-2208-03518}. All these
procedures are integrated into a single solver, implemented in Prolog, which
constitutes the core part of the \setlog tool. Several in-depth empirical
evaluations provide evidence that \setlog is able to solve non-trivial problems
\cite{DBLP:journals/jar/CristiaR20,DBLP:conf/RelMiCS/CristiaR18,DBLP:journals/jar/CristiaR21a,CristiaRossiSEFM13,DBLP:journals/corr/abs-2112-15147},
including the security domain
\cite{DBLP:journals/jar/CristiaR21,DBLP:journals/jar/CristiaR21b}.

\setlog provides constraints encoding most of the set and relational operators used in set-based specification languages such as B \cite{Abrial00} and Z \cite{Spivey00}. 
For example: $\Cup(A,B,C)$
is a constraint interpreted as $C = A \cup B$;
$\In$ is interpreted as set membership (i.e., $\in$); $\Dom(F,D)$ corresponds to $\dom F = D$; $\Subseteq(A,B)$ 
to $A \subseteq B$; $\Comp(R,S,T)$ to $T = R \comp S$ (i.e.,
relational composition); $\ApplyTo(F,X,Y)$ is a weak form of function application; and $\Pfun(F)$ constrains $F$ to be a (partial) function.
The language supports some forms of sets such as the empty set ($\{\}$) and extensional sets ($\{t / A\}$, interpreted as $\{t\} \cup A$); more in Appendix \ref{language}.
Formulas in \setlog are built in the usual way by using the propositional
connectives (e.g., \verb+&+, $\Or$, $\Neg$, $\Implies$), and restricted universal and existential quantifiers ($\Forall$ and $\Exists$, see Section
\ref{rq}).

\setlog can be used as both a programming language and a satisfiability solver. Within certain limits, \setlog code enjoys the \emph{formula-program duality}.

\begin{example}\label{ex:usesetlog}
The following \setlog predicate computes the maximum of a set:
\begin{Verbatim}[numbersep=5pt,
                 numberblanklines=false,
                 fontsize=\small,
                 fontshape=tt,
                 frame=single,rulecolor=\color{white},framesep=2mm]
dec_p_type(smax(set(int),int)).
smax(S,Max) :- Max in S & foreach(X in S, X =< Max & dec(X,int)).
\end{Verbatim}
The first line declares the type of $\mathtt{smax}$: the first argument is a set of integers whereas the second is an integer (more in Appendix \ref{types}). The second line defines a clause (as in Prolog) whose body is a \setlog formula making use of a restricted universal quantifier (RUQ). \verb+dec+ declares the type of \verb+X+.

$\mathtt{smax}$ is a program, so we can execute it:
\begin{Verbatim}[numbersep=5pt,
                 numberblanklines=false,
                 fontsize=\small,
                 fontshape=tt,
                 frame=single,rulecolor=\color{white},framesep=2mm]
{log}=> smax({3,5,1,7,4},Max).
Max = 7
\end{Verbatim}
But $\mathtt{smax}$ is also a formula, so we can prove properties true of it by proving that their negations are unsatisfiable:
\begin{Verbatim}[numbersep=5pt,
                 numberblanklines=false,
                 fontsize=\small,
                 fontshape=tt,
                 frame=single,rulecolor=\color{white},framesep=2mm]
{log}=> neg(smax(S,M) & Y =< M & smax({Y / S},K) implies M = K).
false
\end{Verbatim}
That is, \setlog failed to find a finite set \verb+S+ and integer numbers \verb+Y+, \verb+M+ and \verb+K+ satisfying the above formula. Furthermore, if we attempt to prove an unvalid property \setlog returns a counterexample:
\begin{Verbatim}[numbersep=5pt,
                 numberblanklines=false,
                 fontsize=\small,
                 fontshape=tt,
                 frame=single,rulecolor=\color{white},framesep=2mm]
{log}=> neg(smax(S,M) & smax({Y / S},K) implies M = K).
S = {M/N}, K = Y
Constraint: foreach(X in N,M>=X), Y>=M, foreach(X in N,Y>=X), M neq Y
\end{Verbatim}
\end{example}

In \setlog we do not need an specification and a program; the same piece of code is both the program and its specification. This duality is not the result of integrating two tools but a consequence of the mathematical and computational models behind \setlog.
However, the formula-program duality pays the price of reduced efficiency. For this reason a \setlog program must be considered as a prototype of a real implementation.

\section{\label{encoding}Encoding the Coq Model in \setlog}

In this section we show how different elements of the Coq model described in Section \ref{coq} are encoded in the \setlog language. The complete \setlog code of the Android permissions system is publicly available\footnote{\url{https://www.clpset.unipr.it/SETLOG/APPLICATIONS/android.zip}}. 

Figure \ref{f:grantAutoSL} shows the \setlog encoding of the Coq code shown in Figure \ref{f:grantAutoCoq}.
Differently from the Coq code, the \setlog code gathers all the three Coq definitions in a single \setlog clause (\verb+grantAuto+). 
This is so, for example, because the pre- and post-condition may access the same component of the state or may use the same external function. As \setlog performs a sort of symbolic execution over this code, these double accesses or calls imply more computations thus making automated proof to take considerably longer. These differences between proof assistants and automated provers should be considered when the model is written for either of both. We will further discuss this in Section \ref{relwork}. Below we explain the encoding in more detail.

\begin{figure}
\hspace{3mm}
\begin{Verbatim}[gobble=2,
                 numbers=left,
                 numbersep=5pt,
                 numberblanklines=false,
                 xleftmargin=5mm,fontsize=\small]
  dec_p_type(grantAuto(set(perm),system,perm,idApp,system)).
\\grantAuto(SystemPerm,S,P,A,S_) :-
    state(S,St) & state(S_,St_) &
    perms(St,Perms) & grantedPermGroups(St,MG) & perms(St_,Perms_) & 

    isManifestOfApp(A,M,S) &                     % pre_grantAuto
    use(M,PM) & P in PM & 
    (P in SystemPerm or usrDefPerm(P,S)) &
    (comp({[A,A]},Perms,{}) & PS = {} or applyTo(Perms,A,PS) & P nin PS) & 
    pl(P,dangerous) &
    maybeGrp(P,grp(G)) & applyTo(MG,A,SG) & G in SG &

    foplus(Perms,A,{P/PS},Perms_) &              % post_grantAuto
    updateGrantAuto(S,Perms_,S_).
\end{Verbatim}
\caption{\label{f:grantAutoSL}\setlog encoding of the Coq code shown in Figure \ref{f:grantAutoCoq}}
\end{figure}

The first line declares the type of \verb+grantAuto+ whose head is given in line 2 and whose body is given from line 3 on. The first parameter of \verb+grantAuto+, \verb+SystemPerm+, is implicit in the Coq code as there is declared as a global parameter. \setlog does not support implicit parameters so they have to be explicitly included in every clause where they are needed. Then, $s$ corresponds to \verb+S+, $p$ to \verb+P+, $app$ to \verb+A+ and $s'$ to \verb+S_+ (in \setlog variables begin with a capital letter). For example, the \verb+dec_p_type+ declaration states that the type of \verb+S+ is \verb+system+ which is defined as a tuple. In this case, each field in the Coq record corresponds to a component in the tuple, as follows (see Figure \ref{f:coq:state}):
\begin{Verbatim}[fontsize=\small,frame=single,rulecolor=\color{white},framesep=2mm]
[set(idApp),                       % apps
 set(idApp),                       % alreadyVerified
 rel(idApp,set(idGrp)),            % grantedPermGroups
 .........]
\end{Verbatim}
In order to access each component, we have defined predicates, named  as the Coq fields, that use Prolog unification:
\begin{Verbatim}[fontsize=\small,frame=single,rulecolor=\color{white},framesep=2mm]
dec_p_type(apps(state,set(idApp))).
apps(S,Apps) :-  S = [Apps,_2,_3,_4,_5,_6,_7,_8,_9].
\end{Verbatim}

More importantly, the type of each component of \verb+system+ is different from (although equivalent, in some sense, to) the corresponding Coq type. For instance, the Coq type for \verb+apps+ is \verb+list idApp+, where \verb+idApp+ is of type \verb+Set+. In turn, in \setlog, the type is \verb+set(idApp)+, where \verb+idApp+ is a basic type. According to Coq's and \setlog's semantics \verb+idApp+ means roughly the same in both systems: it is just a set of elements. Conversely, the semantics of \verb+list idApp+ and \verb+set(idApp)+ is quite different because the former means that \verb+apps+ is a list in Coq whereas the latter implies that it is a finite set in \setlog. Recall that in Coq \verb+apps+ is declared as a list although the model uses it as a set. 

Similarly, the Coq type of \verb+grantedPermGroups+ is 
\verb+mapping (list idGrp)+, where \verb+mapping : Set := list item+ and \verb+item+ is the record \verb+{idx:index; inf:info}+. That is, \verb+grantedPermGroups+ is basically a list of ordered pairs. However, as the name suggests, the indented usage is for \verb+grantedPermGroups+ to be a finite map or partial function. Furthermore, the range of the map is supposed to be composed of sets rather than lists. The Coq model includes a state invariant stating that this component should be a partial function (recall Figure \ref{f:coq:allMapsCorrect}). Instead, the \setlog type of \verb+grantedPermGroups+ is \verb+rel(idApp,set(idGrp))+. That is, in \setlog this component is a binary relation whose range are sets. Following the Coq model, the \setlog model includes a similar state invariant constraining \verb+grantedPermGroups+ to be a partial function (see Section \ref{lemmainv}).

Returning to Figure \ref{f:grantAutoSL}, lines 3 and 4 extract from \verb+S+ and \verb+S_+ the system components that are needed in the clause. Differently from Coq, in \setlog we need new variables to access each component. For example, in Coq we can do \verb+(state s)+ to access the state of the system whereas in \setlog we do \verb+state(S,St)+ where \verb+St+ is a new variable and then we use \verb+St+ to access the components of the state.

In line 5 we call \verb+isManifestOfApp(A,M,S)+ where \verb+M+ is a new variable implicitly existentially quantified. That is, the existential quantifier \verb+(exists m:Manifest,...)+ in line 2 of Figure \ref{f:grantAutoCoq} is not necessary in \setlog. The code in lines 6 and 7 corresponds to the Coq code \verb+(isSystemPerm p \/ usrDefPerm p s)+. In Coq \verb+isSystemPerm+ has type \verb+Perm -> Prop+ which in \setlog is encoded as set membership to \verb+SystemPerm+.

Line 8 is interesting because it shows how some complex Coq predicates are encoded as \setlog constraints. Indeed, line 8 corresponds to \texttt{~(exists lPerm:list Perm,...)} in lines 5-6 of Figure \ref{f:grantAutoCoq}. That existential quantifier states that: i) \verb+a+ does not belong to the domain of \verb+(perms (state s))+; or ii) if it does belong then \verb+p+ is not in its image. In \setlog i) is encoded as \verb+comp({[A,A]},Perms,{})+ where \verb+comp(R,S,T)+ is a constraint interpreted as $T = R \comp S$ (where $\comp$ is relational composition), and \verb+[A,A]+ is an ordered pair. In turn, ii) is encoded as \verb+applyTo(Perms,A,PS)+ which is a constraint stating minimum conditions to apply a binary relation to a point. If \verb+Perms+ can be applied to \verb+A+, then it has some image (\verb+PS+); and when \verb+Perms+ cannot be applied to \verb+A+, we take \verb+PS+ as the empty set. See that \verb+PS+ is used in line 11 to set the  new state of the system. 

Encoding predicates such as \verb+~(exists lPerm:list Perm,...)+ as \setlog constraints is not the same than putting those predicates behind Coq definitions. Indeed, \setlog implements those constraints at the semantic level. Therefore, encoding complex predicates as constraints is not just a matter of increasing the readability of the model but of increasing \setlog's automated proving capabilities.  

This is particularly noticeable in line 11 of Figure \ref{f:grantAutoSL} as it encodes \verb+grantPerm+ of Figure \ref{f:grantAutoCoq}. \setlog defines \verb+foplus+ as follows:
\begin{equation}\label{eq:foplus}
 \begin{aligned}
  &\verb|foplus(F,X,Y,G) :-|\\[-1mm]
  &\verb|  F = {[X,Z]/H} & comp() & G = {[X,Y]/H}|\\[-1mm]
  &\verb|  or| \\[-1mm]
  &\verb|  comp({[X,X]},F,{}) & G = {[X,Y]/F}.|
 \end{aligned}
\end{equation}
That is, \verb+G+ is equal to \verb+F+ except in \verb+X+ where: if \verb+X+ is in the domain of \verb+F+ then \verb+Y+ becomes the new image of \verb+X+; if not, \verb+[X,Y]+ is added. When \verb+F+ is a function then only one pair is changed.

Hence, line 11 is equivalent to state that \verb+(perms (state s))+ is equal to \texttt{(perms (state s'))} except in \verb+app+ where either \verb+(app,{p})+ is added to \texttt{(perms (state s))} or the new image of \verb+app+ is its old image plus \verb+{p}+. All this, in turn, is expressed in Coq  with the quantified predicates of \verb+grantPerm+.

In the last line, \verb+updateGrantAuto+ updates the system as follows:
\begin{Verbatim}[fontsize=\small,frame=single,rulecolor=\color{white},framesep=2mm]
updateGrantAuto(S,P,S_) :-
  S = [St,_S2] & St = [_1,_2,_3,_4,_5,_6,_7,_8,_9] &
  S_ = [St_,_S2] & St_ = [_1,_2,_3,P,_5,_6,_7,_8,_9].  
\end{Verbatim}
That is, lines 11-12 of Figure \ref{f:grantAutoSL} correspond to \verb+post_grantAuto+ in Figure \ref{f:grantAutoCoq}.

\begin{remark}
The \setlog encoding of Android enjoys the formula-program duality, as explained in Section \ref{setlog}. That is, on one hand, clauses such as \verb+grantAuto+ can be executed, thus turning them into a sort of prototype API that can be used to program security-related scenarios. On the other hand, the same \setlog code is an specification of the Android permissions system. As such, it is possible to use \setlog to (automatically) prove properties true of the specification. This is the subject of the next section.
\end{remark}

\section{\label{prototype}Encoding Security Properties in \setlog}
As we have explained in Sections \ref{sec:model:stateI} and \ref{coq:properties} the Coq model verifies several properties. These properties can be divided into two classes:  invariance lemmas concerning the concept of valid state (Section \ref{sec:model:stateI}), and security properties (Section \ref{coq:properties}). In the introduction we set as one of the main goals of this paper  assessing \setlog as an automated prover for the kind of properties used to validate the Coq model---having as a long term goal integrating \setlog and Coq to optimize the proof process for certain classes of problems.

Hence, in this section we show how the properties proved true of the Coq model can be encoded as \setlog satisfiability queries. Then, these queries can be executed against the \setlog program described in Section \ref{encoding}. Every time we get a \verb+false+ answer we know that the query is unsatisfiable meaning that its negation is a theorem (see Example \ref{ex:usesetlog}). If, on the contrary, we get a solution (counterexample) we know the query is satisfiable meaning that its negation is not a theorem (property) derivable from the program.

\subsection{\label{lemmainv}Valid State}
In this section we show how two of the properties defining the set of valid states are encoded in \setlog. We start with \verb+allMapsCorrect+ defined in Figure \ref{f:coq:allMapsCorrect}. As can be seen, this predicate is the conjunction of several \verb+map_correct+ predicates. In turn, \verb+map_correct+ states that its argument is a partial function. In \setlog $\texttt{pfun}(F)$ constrains $F$ to be a partial function. Hence, when in Coq we have \texttt{map\_correct (grantedPermGroups (state s))} in \setlog we write\footnote{Some \texttt{dec} predicates are avoided for readability.}:
\begin{Verbatim}[fontsize=\small,frame=single,rulecolor=\color{white},framesep=2mm]
dec_p_type(allMapsCorrect4(system)).
allMapsCorrect4(S) :- 
  dec(GR,rel(idApp,set(idGrp))) & 
  state(S,St) & grantedPermGroups(St,GR) & pfun(GR).
\end{Verbatim}
Note that in \setlog, as well as in Coq, we use a combination of types and constraints to state the desired property. That is, \verb+dec(GR,rel(idApp,set(idGrp)))+ restricts the domain of \verb+GR+ to \verb+idApp+ and the range to \verb+set(idGrp)+, whereas \verb+pfun(GR)+ constrains \verb+GR+ to be a function. The \verb+dec+ predicate is enforced during type checking (cf. Appendix \ref{types}) whereas the \verb+pfun+ constraint is enforced during constraint solving (cf. Appendix \ref{constraintsolving}).

Now we turn our attention to predicate \verb+notDupPerm+ shown in Figure \ref{f:coq:notDupPerm}. As can be seen, this predicate begins with a universal quantification over several variables of different types. The Coq predicate then goes to restrict the quantified variables to belong to different sets\footnote{In Coq they are, formally, lists.}. For instance, \verb+a+ and \verb+l+ must verify \verb+defPermsForApp a l+. In turn, \verb+defPermsForApp+ is divided into two cases; let's analyze the first one. In this case \verb+l+ is the image of \verb+a+ through \verb+(defPerms (environment s))+. Then, in a set-based notation such as \setlog this is equivalent to:
\begin{Verbatim}[fontsize=\small,frame=single,rulecolor=\color{white},framesep=2mm]
environment(S,E) & defPerms(E,DP) & [A,L] in DP
\end{Verbatim}
where \verb+S+ corresponds to \verb+s+, and \verb+E+ and \verb+DP+ are new variables. Hence, in \setlog, instead of quantifying over types, we can use a RUQ (cf. Example \ref{ex:usesetlog} and Section \ref{rq}). In other words, we turn Coq's \verb+forall (a:idApp) (l:list Perm),...+ into \verb+foreach([A,L] in DP,...)+. The second case of \verb+defPermsForApp+ can be treated in a similar way because in this case \verb+a+ and \verb+l+ are components of \verb+sysapp+ which in turn is an element of the set \verb+(systemImage (environment s))+. The quantification over \verb+p+ and \verb+p'+ will be discussed shortly.

Therefore, as a general rule, we turn Coq's universal quantifications into \setlog's RUQ.

Finally, in order to encode \verb+notDupPerm+ in \setlog we define three clauses corresponding to the result of distributing the two cases of \verb+defPermsForApp+ into the rest of the predicate. That is, in \setlog, we have:
\begin{itemize}
\item \verb+notDupPerm1+ where \verb+a+, \verb+a'+, \verb+l+ and \verb+l+' are quantified over \verb+defPerms+
\item \verb+notDupPerm2+ where \verb+a+, \verb+a'+, \verb+l+ and \verb+l+' are quantified over \verb+systemImage+
\item \verb+notDupPerm3+ where \verb+a+ and \verb+l+ are quantified over \verb+defPerms+, whereas \verb+a'+  and \verb+l+' are quantified over \verb+systemImage+
\end{itemize}
As an example we show \verb+notDupPerm3+:
\begin{Verbatim}[fontsize=\small,frame=single,rulecolor=\color{white},framesep=2mm]
dec_p_type(notDupPerm3(system)).
notDupPerm3(S) :-
  environment(S,E) & defPerms(E,DP) & systemImage(E,SS) &
  foreach([[A1,L1] in DP, A2 in SS],[ID2,L2],
    foreach([P1 in L1, P2 in L2],[IP1,IP2],
      IP1 = IP2 implies P1 = P2 & A1 = ID2,
      idP(P1,IP1) & idP(P2,IP2)
    ),
    idSI(A2,ID2) & defPermsSI(A2,L2)
  ).
\end{Verbatim}
As can be seen, the universal quantification over \verb+p+ and \verb+p'+ present in the Coq model becomes the inner formula of the outermost RUQ in \setlog. For instance, \verb+L1+ quantifies over the range of \verb+DP+ and then \verb+P1+ quantifies over \verb+L1+.

As explained in Section \ref{sec:model:stateI}, the definition of valid state is given in terms of a state invariant (see Lemma \ref{lemma:valid-state-correct}). In \setlog, instead of proving that the encoding of $valid\_state$ is a state invariant, we prove that each of the properties included in it is a state invariant. In other words, we prove the encoding in \setlog of:
\begin{gather}
I_1(s) \land s\step{a}s' \rightarrow I_1(s') \notag \\
\dots\dots\dots\dots\dots\dots\dots\dots \label{invsl} \\
I_n(s) \land s\step{a}s' \rightarrow I_n(s') \notag
\end{gather}
for all $a:\Action$ where $valid\_state(s) \defs I_1(s) \land \dots \land I_n(s)$.
This is so because attempting to prove Lemma \ref{lemma:valid-state-correct} may make \setlog to incur in a lengthy computation due to the presence of unnecessary hypothesis. For example, if in order to prove $I_k(s')$ the only necessary hypothesis are $I_k(s)$ and $a$, the presence of $I_j(s)$ ($j \neq k$) may make \setlog to attempt to prove the lemma by \emph{first} exploring $I_j(s)$ instead of $I_k(s)$. Once this proof path is exhausted, \setlog will attempt the proof by exploring $I_k(s)$, which will eventually succeed. The net result is \setlog taking longer than necessary.

Each of the proof obligations in \eqref{invsl} is encoded in \setlog as the following query:
\begin{Verbatim}[fontsize=\small,frame=single,rulecolor=\color{white},framesep=2mm]
neg(I_k(S) & a(S,S_) implies I_k(S_))
\end{Verbatim}
where \verb+I_k+, \verb+a(S,S_)+ and \verb+S_+ correspond to $I_k$, $s\step{a}s'$ and $s'$, respectively.

\subsection{\label{lemmasecprop}Security Properties}
In order to encode in \setlog the security properties described in Section \ref{coq:properties}, we classified them as follows:
\begin{enumerate}
\renewcommand{\theenumi}{\Alph{enumi}}
\item\label{i:p1} Properties given in terms of a sequence of operations. 
\item\label{i:p2} Existential properties, i.e. properties establishing the existence of a state where some predicate holds. 
\item\label{i:p3} Universal properties, i.e. classical system properties. 
\end{enumerate}

In general, properties in class \ref{i:p1} cannot be expressed in \setlog. However, the proof of some of these properties can be divided into two steps: firstly, some predicate is proved to be invariant w.r.t. every operation; secondly, the first result is used to prove that any sequence of operations verifies the desired property. The first step is hard, whereas the second is rather simple. \setlog can be used in the first step by proving that the involved predicate is indeed an invariant (as shown in Section \ref{lemmainv}). Once all these proof obligations have been discharged they could be used in the second step. Hence, we encoded in \setlog the first step and used \setlog to discharge the corresponding invariance lemmas.

A typical property in class \ref{i:p2} is \verb+ExecAutoGrantWithoutIndividualPerms+, shown in Figure \ref{f:execAutoGrantWithoutIndividualPerms}. Its \setlog encoding is the following:
\begin{Verbatim}[
  fontsize=\small,
  frame=single,rulecolor=\color{white},framesep=2mm,commandchars=\\\{\}]
execAutoGrantWithoutIndividualPerms :-
  state(S,St) & perms(St,Perms) &
  validstate(SystemPerm,S) &
  applyTo(Perms,A,PermsA) &
  neg(exists(P_ in PermsA,[MG],MG = grp(G),maybeGrp(P_,MG))) & 
  pl(P,dangerous) &
  maybeGrp(P,grp(G)) &
  pre_grantAuto(SystemPerm,P,A,S).
\end{Verbatim}
As can be seen, the existential quantification of Figure \ref{f:execAutoGrantWithoutIndividualPerms} is implicit in \setlog. Instead of quantifying over \verb+p'+ and \verb+permsA+ we first get \verb+PermsA+ as the image of \verb+A+ through \verb+Perms+; if \verb+A+ is not in \verb+Perms+'s domain, \verb+applyTo(Perms,A,PermsA)+ fails. Then, we only need \verb+P_+ to quantify over \verb+PermsA+ in the \verb+exists+ constraint. In Coq, \verb+Some g+ is part of an \emph{option type} which in \setlog is encoded as \verb+grp(G)+, where \verb+grp+ is a functor and \verb+G+ an existential variable. 

When \verb+execAutoGrantWithoutIndividualPerms+ is executed \setlog attempts to find values for the variables satisfying the formula. In this way \setlog generates a witness (solution) satisfying the query. The following is a simplified, pretty-printed form of that witness.
\begin{Verbatim}[fontsize=\small,frame=single,rulecolor=\color{white},framesep=2mm]
P = [V9,grp(G),dangerous],  
SystemPerm = {P/Q9},
S = [St,Env],
St = [[apps,{A/Q1}],V1,[grantedPermGroups,{[A,{G/Q2}]/Q3}],
      [perms,{[A,PermsA]/Q4}],...],
Env = [[manifest,{[A,[V2,V2,V3,{P/Q5},V4,V5]]/Q6}],[cert,{[A,V6]/Q7}],
       [defPerms,{[A,V7]/Q8}],V8]
\end{Verbatim}
Variables \verb+V?+ and \verb+Q?+ are new (existential) variables. This witness is followed by a number of constraints restricting the values the variables at the right-hand side can take. For example, one such constraint is \verb+comp({[A,A]},Q6,{})+ meaning that \verb+A+ is not in the domain of \verb+Q6+, which is the `rest' of \verb+manifest+. This ensures that \verb+manifest+ is a partial function, as required by \verb+validstate+. In order to come out with a more concrete witness every \verb+Q?+ variable can be replaced by the empty set. For instance, a more concrete value for \verb+manifest+ can be obtained by replacing \verb+Q5+ and \verb+Q6+ with \verb+{}+ thus getting \verb+{[A,[V2,V2,V3,{P},V4,V5]]}+. \setlog guarantees that the constraints following a witness are always satisfied by such a simple replacement because those constraints are of an special kind called \emph{irreducible constraints}
\cite{Dovier00,DBLP:journals/jar/CristiaR20,DBLP:journals/jar/CristiaR21a,cristia_rossi_2021,DBLP:journals/corr/abs-2105-03005}.

Finally, \verb+CannotAutoGrantWithoutGroup+, shown in Figure \ref{f:cannotAutoGrantWithoutGroup}, is a typical property of class \ref{i:p3}. The following is its \setlog encoding.
\begin{Verbatim}[fontsize=\small,frame=single,rulecolor=\color{white},framesep=2mm]
cannotAutoGrantWithoutGroup :-
  neg(
    state(S,St) & grantedPermGroups(St,GR) &
    pl(P,dangerous) &
    maybeGrp(P,grp(G)) & 
    applyTo(GR,A,GA) & 
    G nin GA
    implies neg(grantAuto(SystemPerm,P,A,S,S_))
  ).
\end{Verbatim}
Clearly, the Coq theorem based on proving a universal property becomes a \setlog query attempting to satisfy the negation of that property. The only issue worth to be noted in this encoding is that \verb+exec s (grantAuto p a) s' ok+ is implemented by simply calling \verb+grantAuto+. If the outcome of \verb+exec+ is \verb+ok+ it means that \verb+grantAuto+ could be executed which in \setlog corresponds to \verb+grantAuto+ being satisfiable. 

In this case when \verb+cannotAutoGrantWithoutGroup+ is executed \setlog is unable to find values satisfying the formula so it returns \verb+false+. If, for some reason, there would have been an error in \verb+grantAuto+ such that \verb+cannotAutoGrantWithoutGroup+ would not hold, then \setlog would return a counterexample when the query is executed. The counterexample would help to find the error.

As a summary, it is reasonable to say that the \setlog code is a straightforward, set-based encoding of the Coq model.

\begin{remark}[An Automatically Verified Prototype]
Once all queries representing properties are successfully run, the \setlog program can be regarded as an automatically verified (or certified or correct) prototype of the Android permissions system w.r.t. the proven properties. As we will see in Section \ref{quantrep}, \setlog is able to automatically discharge in a reasonable time 24 out of 27 ($\approx 90\%$) of the properties proposed in the Coq model.
\end{remark}

\subsection{\label{quantrep}Empirical Evaluation}
In this section we provide detailed data on the empirical evaluation resulting after executing all the \setlog queries representing valid state and security properties, described above. The goal of the empirical evaluation is to measure the number of properties that \setlog is able to prove and the amount of time it spends on that. The \setlog code and instructions on how to reproduce this empirical evaluation can be found online\footnote{\url{https://www.clpset.unipr.it/SETLOG/APPLICATIONS/android.zip}}. 

Table \ref{t:experiments} summarizes the results of the empirical evaluation and provides some figures as to compare the \setlog and Coq developments. 
The upper part of the table gives some figures about the size of both projects. 
The Coq development encompasses the model described in Section \ref{coq} and a Coq functional implementation proved to refine the model. In this way we are comparing all the work needed to get an executable prototype in Coq as well as in \setlog.
As can be seen, the \setlog code is larger than the Coq code in terms of LOC but the latter needs forty times more characters than the former (recall \verb+granPerm+). \setlog lines tend to be really shorter---many times only one constraint per line is written. Concerning proofs, Coq needs almost twice as many LOC than \setlog although in terms of characters Coq uses not many more than \setlog. Here, however, it is important to observe that most of the proof text in \setlog can be automatically generated. In effect, 90\% of the proof text in \setlog corresponds to the \emph{statement} of the invariance lemmas that ensure that each operation preserves valid states, and not to the actual proofs.
These statements can be automatically generated by conveniently annotating clauses representing operations and invariants and then calling a verification condition generator which generates lemmas such as \ref{invsl}.
On the other hand, most of the proof text in Coq corresponds to the proof commands written by the user. 
However, these proofs use standard Coq strategies and tactics that do not apply automated methods such as those proposed by Chlipala \cite{Chlipala2013} for engineering large scale formal developments. The use of such methods could reduce the number of lines and the number of tactics used in the proofs.

\begin{table}
\begin{center}
\begin{tabularx}{.856\textwidth}{Xrrrr}
\toprule 
\rowcolor[gray]{.9}
\multicolumn{5}{c}{\textsc{Project Size (specification + proof)}} \\
 & \multicolumn{2}{c}{\textsc{LOC} (K)} 
 & \multicolumn{2}{c}{\textsc{Char} (K)} \\\cmidrule(l){2-3} \cmidrule(l){4-5}
 & \multicolumn{1}{c}{\textsc{Model}}
 & \multicolumn{1}{c}{\textsc{Proofs}} 
 & \multicolumn{1}{c}{\textsc{Model}}
 & \multicolumn{1}{c}{\textsc{Proofs}} \\\cmidrule(l){2-5}
\setlog & 4.2 & 11 &  11 & 448 \\
Coq     & 3.5 & 18 & 154 & 502 \\[3mm]
\rowcolor[gray]{.9}
\multicolumn{5}{c}{\textsc{Performance of Automated Proof (\setlog)}} \\
 & \multicolumn{1}{c}{\textsc{Coq}}
 & \multicolumn{2}{c}{\textsc{\setlog}}
 & \multicolumn{1}{c}{\textsc{Time} (s)} \\\cmidrule(r{2pt}){3-4}
 & 
 & \multicolumn{1}{c}{\textsc{Lemmas}}
 & \multicolumn{1}{c}{\textsc{Queries}}
 & \\\cmidrule(l){2-5}
Valid-state invariance lemmas & 13 & 13 & 756 &   920 \\
Security properties           & 14 & 11 &  45 &   381 \\[2mm]
\textsc{Totals}               & 27 & 24 & 801 & 1,302 \\
\bottomrule
\end{tabularx}
\end{center}
\caption{\label{t:experiments} Summary of the empirical evaluation}
\end{table}
%

In turn, the lower part of the table provides figures about the performance of \setlog concerning automated proof. As we have said in Section \ref{coq}, there are 13 valid-state invariance lemmas and 14 security properties defined and proved in Coq. \setlog is able to automatically discharge all the invariance lemmas and 11 of the security properties. That is 24 out of 27 ($\approx 90\%$) Coq proofs are automatic in \setlog. In order to prove the 13 invariance lemmas we defined 756 \setlog queries. Recall from Section \ref{lemmainv} that some Coq predicates defining the notion of valid state are split in two or more \setlog clauses (e.g. \verb+notDupPerm+). \setlog needs 920 s ($\approx 15$ m) in order to execute all these queries. Same considerations apply to security properties. That is, the 11 properties are encoded as 45 \setlog queries which are discharged in 381 s ($\approx 6$ m). For example, some security properties of class \ref{i:p1} (Section \ref{lemmasecprop}) are expressed as several state invariants. In summary, \setlog needs 1,302 s ($\approx 22$ m) to prove 24 properties expressed as 801 satisfiability queries. 

The above figures were obtained on a
Latitude E7470 (06DC) with a 4 core Intel(R) Core\texttrademark{} i7-6600U CPU
at 2.60GHz with 8 Gb of main memory, running Linux Ubuntu 18.04.6 (LTS) 64-bit
with kernel 4.15.0-184-generic. \setlog 4.9.8-11h over SWI-Prolog
(multi-threaded, 64 bits, version 7.6.4) was used during the experiments.

\section{\label{relwork}Discussion and Related Work}

Besides the Coq model discussed in Section \ref{coq}, De Luca and Luna \cite{DBLP:conf/types/Luca020} developed a Coq functional implementation. Then, they demonstrated, using Coq, that this implementation satisfies the model.
Finally, thanks to the program extraction mechanism offered by Coq, they 
obtained a Haskell implementation of the Android permission system  satisfying the model. This is the way a verified prototype is usually generated in Coq. Conversely, in \setlog it is enough to write the specification and then to prove properties to gain confidence on its correctness.

As the empirical evidence shows, combining the automated proof capabilities of \setlog with the proving power of Coq would be beneficial in terms of human effort and development time. \setlog is able to automatically discharge all but three of the proposed properties, thus saving a lot of time and human effort; these three properties are manually proved in Coq. Hence, the question is which is the better integration strategy to make \setlog and Coq to work together. Traditionally, automated theorem provers (ATP) have been integrated \emph{into} interactive theorem provers (ITP) such as Coq. For example, CoqHammer \cite{DBLP:journals/jar/CzajkaK18} uses external ATPs to automate Coq proofs. Likewise, Isabelle/HOL uses Sledgehammer extended with SMT solvers \cite{DBLP:journals/jar/BlanchetteBP13}. In these cases, the current goal in the ITP is encoded in the input languages of the external solvers. 

We instead propose to integrate \setlog with Coq in the other way around: encoding \setlog models as Coq specifications\footnote{In this work the Coq model was encoded in \setlog simply because our collaboration started when the Coq model was already available.}. That is, we envision the following methodology:
\begin{enumerate}
\item Write a \setlog model.
\item Prove as many properties as possible using \setlog.
\item If there are unproven properties, encode the \setlog model as a Coq specification.
\item Use Coq to prove the remaining properties.
\end{enumerate}
In this way, the \setlog model becomes a correct-by-construction prototype. This implies that it would not be necessary to write a Coq abstract model, its implementation and perform the refinement proofs in order to get a certified executable prototype.

The justification for encoding \setlog models as Coq specifications can be sought in Figures \ref{f:grantAutoCoq} and \ref{f:grantAutoSL}. As we have shown, \verb+grantPerm+ is encoded in \setlog as line 11 of Figure \ref{f:grantAutoSL}. Consider for a moment (automatically) translating \verb+grantPerm+ into \setlog in a way that line 11 is obtained. The task seems overwhelming. It might be tempting to consider more literal Coq-to-\setlog translations where the \setlog code would look like much as \verb+grantPerm+. The problem with this approach is that it will severely reduce the automated proving capabilities of \setlog. Indeed, \setlog is able to discharge many proofs because predicates like \verb+grantPerm+ are encoded as simple \setlog constraints. That is, \setlog ``knows'' what to do with \verb+foplus+ but it will not necessarily know what to do if that is written in a different way. On the other hand, (automatically) encoding \setlog constraints (e.g. \verb+foplus+) in Coq is rather easy. Considering the definition of \verb+foplus+ in equation \eqref{eq:foplus}, it is easy to see that it can be encoded in terms of set equality and relational composition which are already defined in Coq. If duly unfolded, the resulting translation would look much as \verb+grantPerm+. Evidently, Coq is more expressive than \setlog making the proposed translation easier than the opposite. Once the \setlog model is encoded as a Coq specification, properties can be proved as usual.

We believe that one of the core constructs that allow \setlog to discharge many proof obligations are RUQ. Other tools  (e.g SMT solvers) provide (unrestricted) universal quantification. Although universal quantification is important to gain expressiveness it is harder to deal with when performing automated proofs. Given that RUQ quantify over a set, a set solver can deal with them in a more controlled and decidable way \cite{DBLP:journals/jar/CristiaR21, DBLP:journals/corr/abs-2208-03518}. At the same time, RUQ provide just the necessary expressiveness when it comes to software specification.

Sooner or later, automated proof hits a computational complexity wall that makes progress extremely difficult. We use \emph{hypothesis minimization} and \emph{predicate delay} to move that wall as far away as possible. Hypothesis minimization consists in calling \setlog to prove a property just with the necessary hypothesis. That is, instead of calling \setlog to prove $p \land q \implies r$ we call it to prove $q \implies r$ if $p$ does not contribute to the proof. Predicate delay instructs \setlog to delay the processing of a given predicate until nothing else can be done. This is achieved by means of \setlog's \verb+delay+ directive. Similar techniques are used, for instance, by SMT solvers. For example Dafny \cite{DBLP:conf/lpar/Leino10} provides the \verb+opaque+ attribute and the \verb+reveal+ directive to selectively hide irrelevant predicates from Z3. Hawblitzel et al. \cite[Section 6.3.2]{DBLP:conf/sosp/HawblitzelHKLPR15} use these constructs and add to Dafny more such constructs in order to tame the complexity of the correctness proofs of distributed systems.

In anyway, some times these techniques are not enough to make the solver to finish in a reasonable time. As we have pointed out, \setlog is unable to prove three properties that were proved in Coq by De Luca and Luna. These proofs require less than 500 LOC of Coq code meaning a reduced manual effort ($\approx 2\%$ of the total Coq proof effort). This is another piece of evidence suggesting that the proposed integration between \setlog and Coq could work in practice.

\setlog has been used in a similar fashion with other systems. It can discharge in 2 seconds all the 60 proof obligations proving the correctness of the Bell-LaPadula security model \cite{BLP1,BLP2} w.r.t. to the security condition and the \mbox{*-property} \cite{DBLP:journals/jar/CristiaR21}. The Tokeneer ID Station (TIS) was proposed by the NSA as a benchmark for the production of secure software. NSA asked Altran UK
to provide an implementation of TIS conforming to Common Criteria EAL5 \cite{cc315}. Altran UK applied its own Correctness by Construction development process to
the TIS software including a Z specification of the user requirements. Cristi\'a and Rossi \cite{DBLP:journals/jar/CristiaR21b} encoded that Z specification in \setlog as well as all the proof obligations set forth by the Altran UK team as 523 proof queries. \setlog is able to prove all the 523 queries in 14 minutes. Boniol and Wiels proposed a real-life, industrial-strength case study, known as the Landing Gear System (LGS) \cite{DBLP:conf/asm/BoniolW14}. Mammar and Laleau \cite{DBLP:journals/sttt/MammarL17} developed an Event-B specification of the LGS on the Rodin platform \cite{Coleman00}. This specification was encoded in \setlog along with all the proof obligations generated by Rodin \cite{DBLP:journals/corr/abs-2112-15147}.  The tool discharges all the 465 proof obligations in less than 5 minutes.

Few works study the permission system of Android by using a formal specification language. An example of this is the work of Shin et. al. \cite{DBLP:conf/socialcom/ShinKFT10},
where they specified an abstract model of the Android permission system very similarly to the way that De Luca and Luna did in their last work. The main
difference, though, is that the work by Shin is based on an older version of Android and was never updated. More examples of Coq being used to
study Android at a formal level are the recent work carried out by El-Zawawy et. al. \cite{DBLP:journals/computing/El-ZawawyFC22} or the CrashSafe tool \cite{DBLP:journals/hcis/KhanUASAKAA18}. These works are focused, thou, on studying Inter-Component Communication (ICC) properties rather than security properties
of the permission system.
 
On the other hand, not everyone chooses Coq to formalize and analyze a model of Android. For instance, Bagheri et. al. \cite{DBLP:journals/fac/BagheriKMJ18} formalized and studied the Android permission
system using Alloy \cite{DBLP:conf/zum/Jackson03}. As another example of a formal language being used to study Android, we can mention
the Terminator tool \cite{DBLP:conf/icse/SadeghiJGBM18} which uses the TLA+ model checker to analyze and identify permission induced threats.
The main difference
between these approaches and a Coq formalization, is that the latter provides stronger guarantees about the safety and security properties of the platform although requires more human effort to identify
potential flaws. Our goal in this work is to bring together the robustness of
the properties that one can get from a Coq model combined with the automation that tools like Alloy or \setlog can provide. However, it should be observed that Alloy can only prove that formulas are satisfiable, whereas \setlog can prove both, satisfiability and unsatisfiability.
There are more works using formal methods to study a specific aspect of Android which not necessarily conduct formal verification \cite{DBLP:conf/esorics/MicinskiFJFC15,DBLP:conf/esorics/FragkakiBJS12,DBLP:journals/isf/TalegaonK21}.  These approaches require a high
amount of human effort, not only for identifying flaws but also to prove any desired property.

\section{\label{concl}Conclusions}

We present a \setlog model from one in Coq that constitutes an automatically verified executable prototype of the Android permissions system. In particular, we show in detail how the Coq model is encoded in the \setlog language and how automated proofs are performed. We conclude that \setlog can be used as an effective automated prover for systems such as the Android permissions system. In this way much of the manual, expert work needed to prove properties in Coq can be avoided.

Detailed data on the empirical evaluation resulting after executing all the proofs in \setlog is provided. The empirical evidence shows that the combination of the automated proof capabilities of \setlog with the proving power of Coq would be beneficial in terms of human effort and development time.

We have not seen other verification efforts combining Coq and \setlog to produce a certified prototype of a critical system.

Finally, we discuss possible ways to integrate Coq and \setlog as to provide a framework featuring automated proof and prototype generation. Advancing in this direction for the analysis of critical systems (in general) is part of our future work.

\section*{Declarations}
\begin{itemize}
\item Funding: no funding.
\item Conflict of interest: no conflicts of interest.
\end{itemize}

\appendix

\section{\label{app:setlog}More Details on \setlog}

In this appendix we provide more details on \setlog for those readers unfamiliar with it. Cristi\'a and Rossi \cite{zbMATH07552282} give a more thorough presentation of \setlog in a user-oriented stile. 

\subsection{\label{language}The \setlog language}

The arguments passed in to the set and relational constraints supported by \setlog are terms called \emph{set terms} or just \emph{sets}. 
In turn, the elements of these sets can be, basically,
any Prolog uninterpreted term, integers numbers, ordered pairs (written $[x,y]$), other set terms, etc. In this way, in \setlog binary relations are just sets of ordered pairs. The following are the set terms available in \setlog:
\begin{itemize}
\item Variables can be set terms.
\item $\{\}$ is the term interpreted as the empty set.
\item $\{t / A\}$, where $A$ is a set term and
$t$ is an element
is called \emph{extensional set} and is interpreted as
$\{t\} \cup A$. Terms such as $\{t_1,t_2,\dots,t_n / t\}$ and $\{t_1,t_2,\dots,t_n\}$ are also accepted.
\item $\Ris(X \In A, \phi)$, where $\phi$ is any \setlog formula, $A$ is
a any of the above set terms, and $X$ is a bound variable local to the
$\Ris$ term, is called \emph{restricted intensional set} (RIS) and is
interpreted as $\{x : x \in A \land \phi\}$. Actually, RIS have a more complex
and expressive structure \cite{DBLP:journals/jar/CristiaR21a}.
\item $\Cp(A,B)$, where $A$ and $B$ are any set term among the first three, is interpreted
as $A \times B$, i.e., the Cartesian product between $A$ and $B$.
\item $\Int(m,n)$, where $m$ and $n$ are either integer constants or variables,
is interpreted as the integer interval $[m,n]$.
\end{itemize}

With some exceptions any of the above terms can be passed as arguments to the set and relational constraints available in \setlog.

Negation in \setlog has to be treated carefully.
$\Neg$ computes the \emph{propositional} negation of its argument. In
particular, if $\phi$ is an atomic constraint, $\Neg(\phi)$ returns the
corresponding negated constraint. For example, $\Neg(x \In A \And z \Nin C)$
becomes $x \Nin A \Or z \In C$.
However, the result of $\Neg$ is not always correct because, in general, the
negated formula may involve existentially quantified variables, whose negation
calls into play (general) universal quantification that \setlog cannot handle
properly. Hence, there are cases where
\setlog users must manually compute the negation of some formulas.
The same may happen for some logical connectives, such as $\Implies$, whose
implementation uses the predicate $\Neg$: $F \Implies G$ is
implemented in \setlog as $\Neg(F) \Or G$.

\subsection{\label{types}Types in \setlog}
\setlog's type system is thoroughly described elsewhere \cite{DBLP:journals/corr/abs-2205-01713,Rossi00}. Typed
as well as untyped formalisms have advantages and disadvantages
\cite{DBLP:journals/toplas/LamportP99}. For this reason, \setlog users can
activate and deactivate the typechecker according to their needs.

\setlog types are defined according to the following grammar:
\begin{gather*}
\tau ::= \Int \mid \Str \mid Atom \mid \Sumtype([Constr,\dots,Constr]) \mid [\tau,\dots,\tau] \mid \Stype(\tau) \\
Constr ::= Atom \mid Atom(\tau)
\end{gather*}
where $Atom$ is any Prolog atom other than $\Int$ and $\Str$. $\Int$
corresponds to the type of integer numbers; $\Str$ corresponds to the type of
Prolog strings; if $atom \in Atom$, then it defines the type given by the set
$\{atom\mathtt{:}t \mid t \text{ is a Prolog atom}\}$ (these are called \emph{basic types}); $\mathsf{sum}([c_1,\dots,c_n])$, with $2 \leq n$, defines
a sum type\footnote{Tagged union, variant, variant record, choice type, discriminated union, disjoint union, or coproduct.}; $[T_1,\dots,T_n]$ with $2 \leq n$ defines the Cartesian
product of types $T_1,\dots,T_n$; and $\Stype(T)$ defines the powerset type of
type $T$. Besides, \setlog defines: $\Rtype(\tau_1,\tau_2)$ as $\Stype([\tau_1,\tau_2])$ (i.e. the type of all binary relations); and $\Etype([c_1,\dots,c_n])$ as $\Sumtype([c_1,\dots,c_n])$ where all $c_i$ are nullary terms (i.e. enumerated types). This type system is inspired in Z's.

When in typechecking mode, all variables and user-defined predicates must be
declared to be of a precise type. Variables are declared by means of the
$\Dec(V,\tau)$ predicate, meaning that variable $V$ is of type $\tau$. In this
mode, every \setlog atomic constraint has a polymorphic type much as in the Z
notation. For example, in $\Comp(R,S,T)$ the type of the arguments must be
$\Rtype(\tau_1,\tau_2)$, $\Rtype(\tau_2,\tau_3)$ and
$\Rtype(\tau_1,\tau_3)$, for some types $\tau_1, \tau_2, \tau_3$.

Concerning user-defined predicates, if the head of a predicate is
$\texttt{p(}X_1,\dots,X_n\texttt{)}$, then a declaration of the form $\texttt{dec\_p\_type(p(}\tau_1,\dots,\tau_n\texttt{))}$, where $\tau_1, \dots, \tau_n$
are types, must precede \texttt{p}'s definition. This is interpreted by the
typechecker as $X_i$ is of type $\tau_i$ in \texttt{p}, for all $i \in 1 \dots
n$.

This type system permits to encode the types used in the Coq model of the Android permissions system as shown in Section \ref{encoding}.

\subsection{\label{constraintsolving}Constraint solving}
As concerns constraint solving, \setlog is a rewriting system composed of a collection of specialized rewriting procedures. Each rewriting procedure applies a few
non-deterministic rewrite rules to constraints of one kind. At the core of these procedures is set
unification \cite{Dovier2006}. Rewriting finishes either when the resulting formula is $\false$ or when no constraint can be further rewritten. In the last case the resulting formula is returned as part of the computed answer. This formula is guaranteed to be satisfiable. Note that many of these formulas can be returned, due to the presence of non-deterministic rewrite rules. The disjunction of these formulas represents all the
concrete (or ground) \emph{solutions} of the input formula.
 
The collection of rewriting procedures implements various decision procedures for different theories on the
domain of finite sets, finite set relation algebra and integer numbers. Specifically, \setlog implements:
a decision procedure for the theory of \emph{hereditarily finite sets}
($\SET$), i.e., finitely nested sets that are finite at each level of nesting
\cite{Dovier00}; a decision procedure for a very expressive fragment of the
theory of finite set relation algebras ($\BR$)
\cite{DBLP:journals/jar/CristiaR20,DBLP:conf/RelMiCS/CristiaR18}; a decision
procedure for $\SET$ extended with
restricted intensional sets ($\RIS$) \cite{DBLP:journals/jar/CristiaR21a}; a
decision procedure for
$\SET$ extended with cardinality constraints ($\LCARD$)  \cite{cristia_rossi_2021}; a
decision procedure for $\LCARD$ extended with integer intervals ($\LINT$)
\cite{DBLP:journals/corr/abs-2105-03005}; a decision procedure for quantifier-free, decidable languages extended with restricted quantifiers ($\mathcal{RQ}$) \cite{DBLP:journals/corr/abs-2208-03518}; and integrates an existing decision
procedure for the theory of linear integer arithmetic ($\LIA$) \cite{holzbaur1995ofai}.

\subsection{\label{rq}Restricted quantifiers}
As we have said, \setlog supports restricted quantifiers (RQ) called $\Forall$ and $\Exists$. RQ have been used across the \setlog encoding of the Coq model of the Android permissions system and they have a rather complex structure. For these reasons here we present $\Forall$ with more detail---$\Exists$ is used in the same way. $\Forall$ is a \emph{restricted universal quantifier} (RUQ). The most general form of $\Forall$ is the following:
\begin{equation}\label{eq:ruq4}
\Forall(x \In A,[e_1,\dots,e_n],\phi(x,e_1,\dots,e_n),\psi(x,e_1,\dots,e_n))
\end{equation}
where $x$ can be a variable or an ordered pair of two distinct variables; $e_1,\dots,e_n$ are variables implicitly existentially quantified inside the $\Forall$; $\phi$ is a \setlog formula; and $\psi$ is a conjunction of so-called \emph{functional predicates}. A predicate $p$ of artity $n+1$ ($0 < n$) is a functional predicate iff for each $x_1,\dots,x_n$ there exists exactly one $y$ such that $p(x_1,\dots,x_n,y)$ holds; $y$ is called the \emph{result} of $p$. 
For instance, $\Cup$, $\Dom$ and $\Comp$ are functional predicates. In a $\Forall$, $e_1,\dots,e_n$ must be the results of the functional predicates in $\psi$.
The second and last parameters are optional.
The semantics of \eqref{eq:ruq4} is:
\begin{equation*}
\forall x(x \in A \implies (\exists e_1,\dots,e_n (\psi(x,e_1,\dots,e_n) \land \phi(x,e_1,\dots,e_n))))
\end{equation*}

The structure of the $\Forall$ predicate was designed to avoid the introduction of as many `uncontrolled' existential variables as possible. The introduction of existential variables brings in the problem of negating the predicate in such a way that the result is a formula laying inside the decidable fragment. This structure proved to be expressive enough as to work with real-world problems \cite{DBLP:journals/jar/CristiaR21,DBLP:journals/corr/abs-2112-15147,DBLP:journals/corr/abs-2208-03518}.

Nested $\Forall$ can be written as follows:
\begin{equation*}
\Forall([x \In A, y \In B],\phi) \defs \Forall(x \In A, \Forall(y \In B,\phi))
\end{equation*}

\bibliographystyle{sn-mathphys}
\bibliography{/home/mcristia/escritos/biblio}

\end{document}